\documentclass[12pt]{article}
\usepackage{amsfonts, amsmath, amsthm, amssymb, graphicx, latexsym, natbib, titling}
\usepackage[titletoc,title]{appendix}

\usepackage{color} %\color{declared-color}{text}
\usepackage[left=1in,top=1in,right=1in,bottom=1in,nohead,paperwidth=8.5in, paperheight=11in]{geometry} %1 inch margins, 8 1/2 x 11-inch pages
\thanksmarkseries{alph}
\title{Identifying Demand Effects in a Large Network of Product Categories \\\textit{ \small Forthcoming in Journal of Retailing} }
\author{Sarah Gelper\thanks{Innovation, Technology Entrepreneurship \& Marketing Group, Eindhoven University of Technology}, Ines Wilms\thanks{Faculty of Economics and Business, KU Leuven. Corresponding author: ines.wilms@kuleuven.be}, Christophe Croux\thanks{Faculty of Economics and Business, KU Leuven}}

\date{ }

\begin{document}
\maketitle

%\begin{frontmatter}
%
%\author[ndl]{S. Gelper}
%\ead{S.Gelper@tue.nl}
%
%
%\author[be]{I. Wilms}
%\ead{Ines.Wilms@kuleuven.be}
%
%\author[be]{C. Croux}
%\ead{Christophe.Croux@kuleuven.be}
%
%\address[ndl]{Innovation, Technology Entrepreneurship \& Marketing Group, Eindhoven University of Technology}
%\address[be]{Faculty of Economics and Business, KU Leuven}
%
%
%\begin{abstract}
\noindent
\textbf{Abstract.} Planning marketing mix strategies requires retailers to understand within- as well as cross-category demand effects. Most retailers carry products in a large variety of categories, leading to a high number of such demand effects to be estimated. At the same time, we do not expect cross-category effects between all categories. This paper outlines a methodology to estimate a parsimonious product category network without prior constraints on its structure. To do so, sparse estimation of the Vector AutoRegressive Market Response Model is presented. We find that cross-category effects go beyond substitutes and complements, and that categories have asymmetric roles in the product category network. Destination categories are most influential for other product categories, while convenience and occasional categories are most responsive. Routine categories are moderately influential and moderately responsive. \\
%\end{abstract}
%
\bigskip

%\begin{keyword}

\bigskip

%\end{keyword}
%\end{frontmatter}

\newpage
%%%%%%%%%%%%%%%%%%%%%%%%%%%%%%%%%%%%%%%%%%%%%%%%%%%%%%%%%%%%%%%%%%%%%%%%%%%%%%%%%%%%%%%%%
\section{Introduction}
%%%%%%%%%%%%%%%%%%%%%%%%%%%%%%%%%%%%%%%%%%%%%%%%%%%%%%%%%%%%%%%%%%%%%%%%%%%%%%%%%%%%%%%%%

% Why studying cross-category
While within-category demand effects of the marketing mix have been studied extensively, cross-category effects are less well understood \citep{leeflang:12}.  Nevertheless, cross-category effects might be substantial.  Some categories are complements, e.g.\ bacon and eggs studied by \cite{Niraj08} or cake mix and cake frosting studied by \cite{Manchanda99}, while others are substitutes, e.g.\ frozen, refrigerated and shelf-stable juices \citep{Wedel2004}. But cross-effects also exist among categories that are not complements or substitutes for several reasons. First, as a result of brand extensions, brands are no longer limited to one category \citep{Erdem98,Kamakura2007,Ma2012}. So advertising and promotion of a brand within one category might spill over to own brand sales in other categories.
Second, advertising and promotions generate more store traffic and therefore more sales in other categories \citep{Bell1998}. 
And third, lower expenditures in one category alleviate the budget constraint such that consumers are able to spend more on other, seemingly unrelated, categories \citep{song:07, Lee2013}.

While cross-category effects might be substantial for these reasons, we do not expect that each category's marketing mix variables influence each and every other category.  Instead, we expect some cross-category effects to be zero -- or very close to zero -- but we can not a priori exclude them. Therefore, we use an exploratory modeling approach for parsimonious estimation of a product category network. The network allows us to easily identify categories that are influential for or responsive to changes in other categories. Building on a widely used category typology of destination, routine, occasional and convenience categories \citep{Blattberg95,Briesch2013}, we find that destination categories are most influential, convenience and occasional categories most responsive, and routine categories moderately influential and moderately responsive. 

In order to estimate the cross-category network, this paper presents sparse estimation of the Vector AutoRegressive (VAR) model.  The estimation is {\it sparse} in the sense that some of the within-and cross-category effects in the model can be estimated as exactly zero. Initiated by the work of \cite{Baghestani91} and \cite{Dekimpe95}, the VAR Market Response Model has become a standard, flexible tool to measure own- and cross-effects of marketing actions in a competitive environment. The main drawback of the VAR model is the risk of overparametrization because the number of parameters increases quadratically with the number of included categories. 
Earlier studies using the VAR model, like e.g. \citeauthor{Nijs01} (\citeyear{Nijs01}; \citeyear{Nijs2007}); \cite{Pauwels2002}; \citeauthor{Srinivasan00} (\citeyear{Srinivasan00}; \citeyear{Srinivasan04});  \cite{Steenkamp05}, were often limited by this overparametrization problem.
To overcome this problem, previous research on cross-category effects has limited its attention to a small number of categories by studying substitutes or complements \citep{Kamakura2007,song:07,Leeflang:08,Bandyopadhyay2009,Ma2012}.  We present an estimation technique for cross-category effects in much larger product category networks. The technique allows many parameters to be estimated even with short observation periods. Short observation periods are commonplace in marketing practice since many firms discard data that are older than one year \citep{Lodish07}. 

% CONTRIBUTION PAPER
This paper contributes to the extant retail literature in a number of important ways.  
(1) Previous cross-category literature largely limits attention to categories that are directly related through substitution, complementarity or brand extensions. We provide  evidence that cross-category effects go beyond such directly related categories. 
(2) We introduce the concepts of influence and responsiveness of a product category and position different category types (destination, routine, occasional and convenience)  according to these dimensions.  
(3) To identify the cross-category effects, we estimate a large VAR model using an extension of the lasso approach of \cite{Tibshirani96}.

% Paper organization
The remainder of this article is organized as follows. Section 2 positions this paper in the cross-category management literature and describes the conceptual framework that positions category types according to their influence and responsiveness. Section 3 discusses the methodology.  We describe the sparse estimator of the VAR model, discuss how to construct impulse response functions ans compare the sparse estimation technique with two Bayesian estimators. In Section 4, a simulation study shows the excellent performance of the proposed methodology in terms of estimation reliability and prediction accuracy.  Section 5 presents our data and model, Section 6 our findings on cross-category demand effects. 
We first identify which categories are most influential and which are most responsive to changes in other categories. Then, we identify the main cross-category effects based on estimated cross-price, promotion and sales elasticities.

%%%%%%%%%%%%%%%%%%%%%%%%%%%%%%%%%%%%%%%%%%%%%%%%%%%%%%%%%%%%%%%%%%%%%%%%%%%%%%%%%%%%%%%%%
\section{Cross-Category Management}
%%%%%%%%%%%%%%%%%%%%%%%%%%%%%%%%%%%%%%%%%%%%%%%%%%%%%%%%%%%%%%%%%%%%%%%%%%%%%%%%%%%%%%%%%
% Importance of category management for retailers
The importance of category management for retailers is widely acknowledged, both as a marketing tool for category performance \citep{Fader1990, Basuroy2001, Dhar2002} and as an operational tool for planning and logistics \citep{Rajagopalan2012}. 
Successful  category management requires retailers to understand cross-category effects of prices, promotions and sales.
% Price -> Sales
Among these, the cross-category effects of prices on sales  -- which define substitutes and complements -- are the most extensively studied \citep{Song:06, Bandyopadhyay2009, leeflang:12, Sinistyn2012}.  
% Promotion -> Sales
Cross-category effects of promotions, e.g.\ feature and display promotions, on sales result from many brands being active in multiple categories \citep{Erdem02}. Brand associations carry over to products of the same brand in other categories, e.g.\ through umbrella branding \citep{Erdem98} or horizontal product line extensions \citep{Aaker90}. 
% Sales -> Sales 
% - Budget constraint, Destination category
Less well understood than the effects of prices and promotions, are the effects of sales in one category on sales in other categories.  Such effects might exist because categories are related based on affinity in consumption \citep{Shankar2014}, because products from various categories are placed close to each other in the shelves \citep{Bezawada09, Shankar2014}, or because of the budget constraint \citep{Du2008}.  If consumers spend more in a certain category they might, all else equal, spend less in other categories simply because they hit their budget constraint.  As a result, cross-category effects might exist between seemingly unrelated categories. 

% ASYMMETRY
When studying these cross-category effects of price, promotion and sales on sales, several asymmetries might arise. 
% Within versus cross-category effects
A first asymmetry concerns within- versus cross-category effects. We expect within-category effects to be more prevalent and larger in size than cross-category effects (e.g. \citealp{Song:06}; \citealp{Bezawada09}).
% Role of categories
A second asymmetry concerns category influence versus category responsiveness.
%Category influence and responsiveness
Influential categories are important drivers of other category's sales, while sales of responsive categories react to changes in other categories.  To identify which categories are more influential or more responsive, we build on a widely used typology of categories described in \cite{Blattberg95}. 

\cite{Blattberg95} define 4 category types from the consumer perspective: destination, routine, occasional and convenience. 
%Destination
Destination categories contain goods that consumers plan to buy before they go on a shopping trip, such as soft drinks. \cite{Briesch2013} show that destination categories are generally categories in which consumers spend a lot of their budget. Retailers typically use a price aggressive promotion strategy and high promotion intensity for these destination categories with the goal of increasing store traffic.
Because consumers shop to buy products in the destination categories, destination categories are likely to influence sales in other categories. However, since consumers already plan to buy in the destination categories before entering the store, destination category sales will not be highly responsive \citep{Shankar2014}.  

%Routine
About 55\% to 60\% of categories are routine categories \citep{Pradhan09}. Routine categories are regularly and routinely purchased, such as juices and biscuits.  Retailers typically use a consistent pricing strategy and average level of promotion intensity. Because purchases in routine categories can more easily be delayed than purchases in destination categories, we expect routine categories to be more responsive.  But, since purchases in routine categories altogether still account for a large portion of the budget, they are also likely to influence sales in other categories. 

%Occasional
Occasional categories follow a seasonal pattern or are purchased infrequently. These categories comprise a small proportion of retail expenditures while they contain typically more expensive items, like oatmeal.  We therefore expect occasional categories to be less influential and more responsive than destination or routine categories. 

%Convenience
Finally, convenience categories are categories that consumers find convenient to pick up during their one-stop shopping trip, like ready-to-eat-meals. These purchase decisions are typically made in the store. Since convenience categories are geared towards consumer convenience and filling impulse needs, we expect them to be highly responsive.

%%%%%%%%%%%%%%%%%%%%%%%%%%%%%%%%%%%%%%%%%%%%%%%%%%%%%%%%%%%%%%%%%%%%%%%%%%%%%%%%%%%%%%%%%
\section{Sparse Vector Auto-Regressive Modeling}
%%%%%%%%%%%%%%%%%%%%%%%%%%%%%%%%%%%%%%%%%%%%%%%%%%%%%%%%%%%%%%%%%%%%%%%%%%%%%%%%%%%%%%%%%

\subsection{Motivation}

% Research question
The aim of this paper is to identify cross-category demand effects in a large product category network.
% Why we need the VAR model
To this end, we use the Vector AutoRegressive (VAR) model. The VAR is ideal for measuring within- and cross-category effects of marketing actions since it accounts for both inertia in marketing spending and performance feedback effects by treating marketing variables as endogenous \citep{Dekimpe95}. 
Other studies on cross-category effects, like e.g.\ \cite{Wedel2004} use a demand model with exogenous prices, or a simultaneous equations model without lagged effects like \cite{Shankar2014}. However, managers may set marketing instruments strategically in response to market performance and market response expectations. Not accounting for time inertia or feedback effects limits our understanding of how the market functions and misleads managerial insights and prediction.

% Challenge
Identifying cross-category demand effects using VAR analysis remains challenging because the sheer number of such effects makes them hard to estimate. The number of parameters to be estimated in the  VAR rapidly explodes, making standard estimation inaccurate. This undermines the ability to identify important relationships in the data.
% Previous literature and limitations
To overcome an explosion of the number of parameters in the VAR, marketing researchers have used pre-estimation dimension reduction techniques, i.e.\ they first impose restrictions on the model and then estimate the reduced model. Four such common techniques are (i) treating marketing variables as exogenous (e.g. \citealp{Nijs01}; \citealp{Pauwels2002} and \citealp{Nijs2007}), (ii) estimating submodels rather than a full model (e.g. \citealp{Srinivasan00}; \citealp{Srinivasan04}), (iii) aggregating or pooling over, for instance, stores or competitors (e.g. \citealp{Horvath08}; \citealp{Slotegraaf2008}), and (iv) applying Least Squares to a restricted model (e.g. \citealp{Dekimpe95, Dekimpe99_b}; \citealp{Nijs2007}). Most researchers applying  pre-estimation dimension reduction techniques recognize that they do so because of the practical limitations of standard estimation techniques rather than for theoretical reasons (e.g. \citealp{Srinivasan04} and \citealp{Bandyopadhyay2009}).  

% Sparse estimation as an outcome
To address the overparametrization of the VAR, we use sparse estimation.  Sparsity means that some of the within- and cross-category effects in the VAR are estimated as exactly zero. 
As argued in the previous section, from a substantive perspective, we cannot exclude cross-category effects before estimation because cross-category effects might occur between seemingly unrelated categories. 
From a methodological perspective, sparse estimation is a powerful solution to handle the overparametrization of the VAR.
In our cross-category model, we endogenously model sales, promotion and prices of 17 product categories. Hence, already in a VAR model with one lag, as much as $(3 \times 17) \times (3 \times 17) = 2601$ within- and cross-category effects need to be estimated. Since the sparse estimation procedure puts some of these effects to zero, a more parsimonious model is obtained. Results are easier to interpret and, therefore, the sparse estimation procedure provides actionable insights to managers.

\subsection{Extending the Lasso to the VAR model}
In situations where the number of parameters to estimate is large relative to the sample size, the Lasso proposed by \cite{Tibshirani96} provides a solution within the multiple regression model.  The Lasso minimizes the least squares criterion penalized for the sum of the absolute values of the regression parameters.  This penalization forces some of the estimated regression coefficients to be exactly zero, which results in selection of the pertinent variables in the model. The Lasso method is well established \citep{Buhlmann2010, Chatterjee2011} and shows good performance in various applied fields \citep{Wu2009, Fan2011}.

The Lasso technique can not be directly applied to the VAR model because the VAR model differs from a multiple regression model in two important aspects. First, a VAR model contains several equations, corresponding to a multivariate regression model. Correlations between the error terms of the different equations need to be taken into account.
Second, a VAR model is dynamic, containing lagged versions of the same time series as right-hand side variables of the regression equation. Both aspects of VAR models make it necessary to extend the lasso to the VAR context, what the sparse estimator in this paper does.

It builds further on a sparse estimator of the multivariate regression model \citep{Rothman10}, and the groupwise lasso for categorical variables \citep{Yuan06, Meier07}. The estimator is consistent for the unknown model parameters, see \citet{Meier07} and \citet{Friedman07}.

\subsection{Model Specification}
Sales, price and promotion are measured for several categories over a certain time period. We collect all these time series in a multivariate time series ${\bf y}_t$ with $q$ components. In our cross-category demand effects study, ${\bf y}_t$ contains sales, price and promotion for 17 product categories, hence $q= 3 \times 17 = 51$. The VAR Market Response Model is given by
\begin{equation}\label{varp}
{\bf y}_t = B_1 {\bf y}_{t-1}  + B_2 {\bf y}_{t-2}  + \ldots + B_p {\bf y}_{t-p}  + {\bf e}_t \, ,
\end{equation}
where $p$ is the lag length. The autoregressive parameters $B_1$ to $B_p$ are $(q \times q)$ matrices, which capture both within- and cross-category effects. The elements of these matrices measure the effect of sales, price and promotion in one category on the sales, price and promotion in other categories (including its own). The error term ${\bf e}_t$ is assumed to follow a $N_q(0,\Sigma)$ distribution. We assume, without loss of generality, that all time series are mean centered such that no intercept is included.

If the number of components $q$ in the multivariate time series is large,  the number of unknown elements  in the sequence of matrices $B_1,\ldots,B_p$ explodes to $p q^2$, and accurate estimation by standard methods is no longer possible. Sparse estimation, with many elements of the matrices  $B_1,\ldots,B_p$ estimated as zero,  brings an outcome: it will not only provide  estimates with smaller mean squared error, but also substantially improve model interpretability.
The method we propose does not require the researcher to prespecify which entries in the $B_j$ matrices are zero and which are not. Instead, the estimation and variable selection are  simultaneously performed. This is particularly of interest in situations where there is no a priori information on which time series is driving which.

The instantaneous correlations in model \eqref{varp} are captured in the error covariance matrix $\Sigma$. If the dimension $q$ is large relative to the number of observations, estimation of $\Sigma$ becomes problematic. The estimated covariance matrix risks getting singular, i.e.\ its inverse does not exist.  Hence, we also induce sparsity in the estimation of the inverse error covariance matrix  $\Omega=\Sigma^{-1}$. The  elements of $\Omega$ have a natural interpretation as  partial correlations between the error components of the $q$ equations in model \eqref{varp}. If the $ij$-th element of the inverse covariance matrix is zero this means that, conditional on the other error terms, there is no correlation between the error terms of equations $i$ and $j$.

\subsection{Penalized Likelihood Estimation}
This section defines the sparse estimation procedure for the VAR model. The Sparse VAR estimator is defined by minimizing a measure of goodness-of-fit to the data  combined with a {\it penalty} for the magnitude of the model parameters.  It is convenient to first recast model \eqref{varp} in stacked form as
\begin{equation}\label{stacked}
y = X \beta + e \, ,
\end{equation}
where $y$ is a vector of length $n q$ containing the stacked  values of the time series. If the multivariate time series has length $T$, then $n=T-p$ is the number of time points for which all current and lagged observations are available. The vector $\beta$ contains the stacked
vectorized matrices $B_1,\ldots,B_p$, and $e$ the vector of stacked error terms.
The matrix $X=I_q \otimes X_0$, with $ X_0 = (\underline{{\bf Y}}_1, \ldots, \underline{{\bf Y}}_p)$, is of dimension $(n q \times p q^2)$.
Here $\underline{{\bf Y}}_j$ is an $(n \times  q)$ matrix, containing the values of the $q$  series at lag $j$ in its columns,  for $1 \leq j \leq p$, with $p$ the maximum lag. The symbol  $\otimes$ stands for the Kronecker product.

The sparse estimator of the autoregressive parameters $\beta$ and the inverse covariance matrix $\Omega=\Sigma^{-1}$ are obtained by minimizing the negative log likelihood with a groupwise penalization on the $\beta$ and a penalization on the off-diagonal elements of $\Omega$:
\begin{equation}\label{mincrit}
(\hat{\beta},\hat{\Omega}) = \underset{(\beta,\Omega)}{\operatorname{argmin}} \, \frac{1}{n} (y-X \beta)^{\prime} \tilde{\Omega} (y-X \beta) - \log|\Omega|  + \lambda_1 \sum_{g=1}^{G} ||\beta_g|| + \lambda_2 \sum_{k \neq k'} |\Omega_{kk'}| \, ,
\end{equation}
where $||u||= (\sum_{i=1}^{n} u_i^2)^{1/2}$ is the Euclidean norm and $\tilde{\Omega}= \Omega \otimes I_n$. 
By simultaneously estimating $\beta$ and $\Omega$, we take the correlation structure between the error terms into account.
The vector $\beta_g$ in \eqref{mincrit} is a subvector of $\beta$,  containing the  regression coefficients for the  lagged values of  the same time series in one of the $q$ equations in model \eqref{varp}. The coefficients of the lagged values of the same time series form a group. The total number of groups is $G=q^2$ because there are $q$ groups within each of the $q$ equations.
The penalty on the regression coefficients enforces that either \textit{all} elements of the group $\hat\beta_g$ are zero or \textit{none}. As a result, we take the dynamic nature of the VAR model into account since the estimated $B_j$ matrices, for $j=1,\ldots,p$, have their zero elements in exactly the same cells.
The penalization on the off-diagonal elements of $\Omega$ induces sparsity in the estimate $\hat\Omega$.  Finally, the scalars $\lambda_1$ and $\lambda_2$ control the degree of sparsity of the regression estimator and the inverse covariance matrix estimator, respectively. The larger these values, the more sparsity is imposed.
Details on the algorithm  to perform penalized likelihood estimation and the selection of the sparsity parameters $\lambda_1$ and $\lambda_2$ can be found in Appendix A.

% Reference to other Sparse methods
Our approach is similar to \cite{Hsu08} who use the Lasso within a VAR context. However, they do not account for the group-structure in the VAR model, nor do they impose sparsity on the error covariance matrix. 
\cite{Davis12} propose another sparse estimation procedure for the VAR. They infer the sparsity structure of the autoregressive parameters from an estimate of the partial spectral coherence using a two-step procedure. Since variable selection is performed prior to model estimation, the resulting estimator suffers from pre-testing bias. Moreover, the number of parameters might still approach the sample size, leading to unstable estimation or even making estimation infeasible if the number of parameters still exceeds the sample size.
Sparse estimation in economics is a growing field, see \cite{Fan2011} and references therein for an overview. 

\subsection{Alternative: Bayesian Estimators}
An alternative to the sparse estimation technique is to impose prior information in a Bayesian setting. Bayesian regularization techniques have been proposed for the VAR model in \cite{Litterman80} and are used in various applied fields such as macroeconomics \citep{Gefang14, Banbura10}, finance \citep{Carriero12} and marketing \citep{Lenk09,Fok:12,Bandyopadhyay2009}. They are also applicable to a situation like ours where there are many parameters to be estimated with a limited observation period, and are thus a good benchmark.  However, these methods are not sparse, they do not perform variable selection simultaneously with model estimation. The following two paragraphs elaborate on two Bayesian estimators which serve as non-sparse alternatives.
\bigskip

{\it Minnesota Prior.} The original Minnesota prior only specifies a prior distribution for the regression parameters of the VAR model. The  error covariance matrix $\Sigma$ is assumed to be diagonal, and estimated by $\hat{\Sigma}_{ii} = \hat{\sigma}_{i}^{2}$ with $\hat{\sigma}_{i}^{2}$ the standard OLS estimate of the error variance in an AR$(p)$ model for the $i^{th}$ time series \citep{koop:09}. The prior distribution of the regression parameters is taken to be multivariate normal:
\begin{equation}
\beta \sim N(\underline{\beta}_{M},\underline{V}_{M}) \label{Minnesotaprior}.
\end{equation}
For the prior mean, the common choice is $\underline{\beta}_{M}=0_{Kq}$ for stationary series. The prior covariance matrix $\underline{V}_{M}$ is diagonal. The posterior distribution is again multivariate normal. Full technical details can be found in \cite{koop:09}.

The main advantage of the Minnesota prior is its ease of implementation, since posterior inference only involves  the multivariate normal distribution. However, imposing the Minnesota prior only ensures that the parameter estimates are \textit{shrunken} towards zero, while the Sparse VAR ensures that some parameters will be estimated as \textit{exactly} zero.
\bigskip

{\it Normal Inverted Wishart Prior.}
The Minnesota prior takes the error covariance matrix $\Sigma$ as fixed and diagonal and, hence, not as an unknown parameter. To overcome this problem, \cite{Banbura10} impose an inverse Wishart prior on the $\Sigma$ matrix. More precisely,

\begin{equation}
\beta \mid \Sigma \sim N(\underline{\beta}_{NIW},\Sigma \otimes \Omega_{0}) \text{\ \ and  \ \ } \Sigma \sim iW(S_{0},\nu_{0}), \label{LBVArRprior}
\end{equation}
where $\underline{\beta}_{NIW},\Omega_{0},S_{0}$ and $\nu_{0}$ are hyperparameters. Under this normal inverted Wishart prior (labeled in the remainder of this paper as ``NIW"), the posterior for $\beta$, conditional on $\Sigma$ is normal, and the posterior for $\Sigma$ is again inverted Wishart. Full technical details can be found in \cite{Banbura10}.

\subsection{Impulse Response Functions}
Impulse response functions (IRFs) are extensively used to assess the dynamic effect of external shocks to the system such as changes in the marketing mix. An IRF pictures how a change to a certain variable at moment $t$ impacts the value of any other time series at time $t+k$, accounting for interrelations with all other variables. The magnitude of the effect  is plotted as a function of $k$. An extensive discussion on the interpretation of the IRF in marketing modeling can be found in \cite{Dekimpe95}. We use IRFs to gain insight in the dynamics of within and cross-category sales, promotion and price effects on each of the 17 product category sales. The IRFs are easily computed as a function of  the Sparse VAR estimator (see \citealp{Hamilton91}). Since we want to account for correlated error terms, we use generalized IRFs \citep{Pesaran1998, Dekimpe99a}.

To obtain confidence bounds for the generalized IRFs estimated by Sparse VAR, we use a residual parametric bootstrap procedure \citep{Chatterjee2011}. We generate $N_{b} = 1000$ time series of length $ T $ from the VAR model (\ref{stacked}). The invertible estimate of $ \Sigma $ delivered by the Sparse VAR estimation procedure is needed to draw random numbers for the $ N_{q}(0,\Sigma) $ error distribution. For each of these $ N_{b} $ multiple time series, the estimates of the regression parameters are computed. We compute the covariance matrix of the $N_{b}$ bootstrap replicates. For each of the $N_b$ generated series impulse response functions are computed; the 90\% confidence bounds are then obtained by taking the 5\% and 95\% percentiles.

%%%%%%%%%%%%%%%%%%%%%%%%%%%%%%%%%%%%%%%%%%%%%%%%%%%%%%%%%%%%%%%%%%%%%%%%%%%%%%%%%%%%%%%%%
\section{Estimation and prediction performance}
%%%%%%%%%%%%%%%%%%%%%%%%%%%%%%%%%%%%%%%%%%%%%%%%%%%%%%%%%%%%%%%%%%%%%%%%%%%%%%%%%%%%%%%%%
We conduct a simulation study to  compare the proposed Sparse VAR with Bayesian methods using the Minnesota and NIW prior.  As benchmarks, we include the classical Least Squares (LS) estimator and two restricted versions of LS which are often used in practice. In the 1-step Restricted LS \citep{Dekimpe95,Dekimpe99_b}, we estimate the model with classical LS, delete all variables with $|$$t$-statistic$|$  $ \leq 1 $, and re-estimate the model with the remaining variables. We also consider an iterative Restricted LS method described in \cite{Lutkepohl04} where we fit the full model using LS and sequentially eliminate the variables leading to the largest reduction of BIC until no further improvement is possible, of which a close variant was used by \cite{Nijs2007}.

We simulate from a VAR model with $q=10$ dimensions and $p=2$ lags.  Each time series has an own auto-regressive structure and we include system dynamics among the different series.  The first series leads series two to five, while the sixth series leads time series 7 to 10. Specifically, the data generating processes are given by
$${\bf y}_t =
\begin{bmatrix}
B_{1} & 0\\
0 & B_{1}\\
\end{bmatrix} {\bf y}_{t-1}
+
\begin{bmatrix}
B_2 & 0\\
0 & B_2\\
\end{bmatrix} {\bf y}_{t-2}
+ {\bf e}_t \, ,
$$
with
\footnotesize
$$
B_{1} =
\begin{bmatrix}
0.4 & 0.0 & 0.0 & 0.0 & 0.0 \\
0.4 & 0.4 & 0.0 & 0.0 & 0.0 \\
0.4 & 0.0 & 0.4 & 0.0 & 0.0 \\
0.4 & 0.0 & 0.0 & 0.4 & 0.0 \\
0.4 & 0.0 & 0.0 & 0.0 & 0.4 \\
\end{bmatrix} \text{and}
\hspace{0.2cm} \hspace{0.2cm}
B_{2} =
\begin{bmatrix}
0.2 & 0.0 & 0.0 & 0.0 & 0.0 \\
0.2 & 0.2 & 0.0 & 0.0 & 0.0 \\
0.2 & 0.0 & 0.2 & 0.0 & 0.0 \\
0.2 & 0.0 & 0.0 & 0.2 & 0.0 \\
0.2 & 0.0 & 0.0 & 0.0 & 0.2 \\
\end{bmatrix}.
$$
\smallskip
\normalsize

In total, there are $p q^2=200$ regression parameters to be estimated with 36 true parameter values different from zero. The 10-dimensional error term ${\bf e}_t $ is drawn from a multivariate normal with mean zero and covariance matrix $\Sigma=0.1 I_{10}$. We generate $N_s=1000$ multivariate time series of length 50 according to the above simulation scheme.

\subsection{Performance measures}

We evaluate the different estimators in terms of (i) estimation accuracy, (ii) sparsity recognition performance, and (iii) forecast performance.

To evaluate estimation accuracy, we compute the mean absolute estimation error (MAEE), averaged over the simulation runs and over the 200 parameters
$$
\mbox{MAEE} = \frac{1}{N_s} \frac{1}{pq^2} \sum_{s=1}^{N_s} \sum_{j=1}^p \sum_{k,l=1}^q  | \hat{b}^s_{klj} - b_{klj} |,
$$
where $\hat{b}^s_{klj}$ is the estimate of  $b_{klj}$,  the $kl^{th}$\ element of the  matrix $B_j$ corresponding to lag $j$, for the $s^{th}$ simulation run.

Concerning sparsity recognition, we compute the true positive rate and true negative rate
\begin{gather}
\text{TPR}(\hat{b},b) = \dfrac{  \# \{ (k,l,j) :  \hat{b}_{klj} \neq 0 \  and \ b_{klj} \neq 0 \}}{\# \{ (k,l,j) :  \ b_{klj} \neq 0 \}} \nonumber \\
\text{TNR}(\hat{b},b) = \dfrac{  \# \{ (k,l,j) :  \hat{b}_{klj} = 0 \  and \ b_{klj} = 0 \}}{\# \{ (k,l,j) :  \ b_{klj} = 0 \}}. \nonumber
\label{sparsityperformance} 
\end{gather}
The true positive rate (TPR) gives an indication on the number of true relevant regression parameters detected by the estimation procedure. The true negative rate (TNR) measures the hit rate of detecting a true zero regression parameter. Both should be as large as possible.

Finally, we conduct an out-of-sample rolling window forecasting exercise. Using the same simulation design as before, we generate multivariate time series of length $T=60$, and use a rolling window of length $S=50$. For all estimation methods, 1-step-ahead forecasts are computed for $t=S,\ldots,T-1$. Next, we compute the Mean Absolute Forecast Error (MAFE), averaged over all time series and across time
\begin{equation}
\text{MAFE} = \frac{1}{T-S} \frac{1}{q} \sum_{t=S}^{T-1}\sum_{i=1}^{q} | \ \hat{y}^{(i)}_{t+1} -  y^{(i)}_{t+1}  \ | ,
\end{equation}
where $y^{(i)}_{t+1}$ is the value of the  $i^{th}$ time series at time ${t+1}$.

\subsection{Results}
Table \ref{simulationresults} presents the performance measures of the Sparse VAR, the Bayesian and benchmark methods. The Sparse VAR estimator performs best in terms of estimation accuracy. It attains the lowest value of the MAEE (0.041).  A paired $t$-test confirms that the Sparse VAR significantly outperforms the other methods (all $p$-values $<0.001$).

\linespread{1.2}
\begin{table}
\begin{center}
\caption{Mean Absolute Estimation Error (MAEE), True Positive Rate (TPR), True Negative Rate (TNR) and Mean Absolute Forecast Error (MAFE), averaged over 1000 simulation runs, are reported for every method. \label{simulationresults}}
\begin{tabular}{lcccccccccccc}
  \hline
Method  					&&& MAEE &&& TPR &&& TNR &&& MAFE  \\ \hline
Sparse VAR 					&&& 0.041 &&& 0.860 &&& 0.848 &&& 0.359 \\
LS 							&&& 0.157 &&& 1 &&& 0 &&& 0.540 \\
Restricted LS: 1-step  		&&& 0.121 &&& 0.709 &&& 0.541 &&& 0.520 \\
Restricted LS: Iterative  	&&& 0.116 &&& 0.261 &&& 0.775 &&& 0.516 \\
Bayesian: Minnesota 		&&& 0.044 &&& 1 &&& 0 &&& 0.355 \\
Bayesian: NIW 				&&& 0.077 &&& 1 &&& 0 &&& 0.476 \\
\hline
\end{tabular}
\end{center}
\end{table}
\linespread{1.5}

Sparsity recognition performance is evaluated using the true positive rate and the true negative rate, reported in Table \ref{simulationresults}. For the LS and Bayesian estimators, all parameters are estimated as non-zero, resulting in a perfect true positive rate and zero true negative rate. Among the variable selection methods, the Sparse VAR performs best. Sparse VAR achieves a value of the true positive rate of 0.86; 0.85 for the true negative rate.

Finally, we evaluate the forecast performance of the different estimators by the Mean Absolute Forecast Error in Table \ref{simulationresults}. The Sparse VAR and the Bayesian estimator with Minnesota prior achieve the best forecast performance. A Diebold-Mariano test confirms that these two methods perform significantly better than the others ($p$-values $<0.001$). There is no significant difference in forecast performance between Sparse VAR and the Bayesian estimator with Minnesota prior.

\subsection{Robustness checks}
\textit{Alternative penalty function.} We investigate the robustness of Sparse VAR to the choice of the penalty function. We replace the grouplasso penalty on the regression coefficients with the elastic net penalty \citep{Zou05}. Elastic net is a regularized regression method that linearly combines the $L_1$ and $L_2$ penalties of respectively lasso and ridge regression. Like the grouplasso, elastic net produces a sparse estimate of the regression coefficients. All other steps of the methodology remain unchanged.  We find that the grouplasso penalty performs slightly better than the elastic net  penalty in terms of estimation accuracy, sparsity recognition and prediction performance. 

\medskip

\textit{Sensitivity to the order of the VAR.} We estimate the model with Sparse VAR for different values of $p$ and evaluate the performance. As expected, Sparse VAR attains the best estimation accuracy for the true value $p=2$. The results are, however, very robust to the choice of the order of the VAR. Selecting $p$ too low is slightly worse than selecting $p$ too high. 

\medskip

\textit{Sensitivity to the sparsity parameters.} The sparsity parameters are selected according to the BIC and this selection is an integral part of the estimation procedure.
The results are not sensitive to the value of $\lambda_2$, which controls the sparsity of $\widehat{\Omega}$. The results are more sensitive to the choice of $\lambda_1$, since it directly influences the sparsity of the autoregressive parameters. It turms out that Sparse VAR still outperforms the other estimators for a large range of $\lambda_1$ values.

%%%%%%%%%%%%%%%%%%%%%%%%%%%%%%%%%%%%%%%%%%%%%%%%%%%%%%%%%%%%%%%%%%%%%%%%%%%%%%%%%%%%%%%%%
\section{Data and Model}
%%%%%%%%%%%%%%%%%%%%%%%%%%%%%%%%%%%%%%%%%%%%%%%%%%%%%%%%%%%%%%%%%%%%%%%%%%%%%%%%%%%%%%%%%
We use the sparse estimation technique for large VARs described in Section 3 to identify cross-category demand effects across 17 categories in the Dominick's Finer Foods database.  This database is a well-established source of weekly scanner data from a large Midwestern supermarket chain, Dominick's Finer Foods (e.g. \citealp{Kamakura2007, Pauwels07}).  We first describe the data and model in more detail, and then report on the insights the Sparse VAR generates in the next section.

%%--------------------------------------------------------------------------------------------
%\subsection{Data and Model}
%%--------------------------------------------------------------------------------------------

We use all 17 product categories in the Dominick's Finer Foods database containing food and drink items, a much broader selection of categories than previous studies on cross-category demand effects have considered. A description of each product category can be found in Table \ref{Categories}. For 15 stores, we obtain weekly sales, pricing  and promotional feature and display data for the 17 product categories.

\linespread{1.2}
\begin{table}
\small
\begin{center}
\caption{Description of the 17 categories from Dominick's Finer Foods database that are analyzed in this paper. For each category, we report the proportion of food and drink expenditures.  \label{Categories}}
\small
\begin{tabular}{lclllc} \hline
Category & Expenditures &&&  Category & Expenditures \\ \hline
Soft Drinks & 22.24\% 		&&& Snack Crackers & 3.04\%\\
Cereals & 13.92\% 			&&& Frozen Juices & 2.88\% \\
Cheeses & 10.46\% 		&&& Canned Tuna & 2.80\% \\
Refrigerated Juices & 7.36\% 			&&&  Frozen Dinners & 2.00\% \\
Frozen Entrees & 6.98\% 	&&&  Front-end-candies & 2.00\% \\
Beer & 6.35\% 				&&&  Cigarettes & 1.49\% \\
Cookies & 6.21\%				&&&  Oatmeal & 1.43\%\\
Canned Soup & 4.82\% 			&&&  Crackers & 1.37\%\\
Bottled Juices& 4.66\% 		&&& & \\ \hline
\end{tabular} \\
\end{center}
\end{table}
\linespread{1.5}
\normalsize

\noindent
{\bf Sales.} Category sales volumes for the 17 categories, measured in dollars per week.

\noindent
{\bf Promotion.} The promotional data include the percentage of SKUs of each category that are promoted (feature and display) in a given week, following \citet{Srinivasan04}.

\noindent
{\bf Prices.} To aggregate pricing data from the SKU level to the product category level, we follow \citet{Srinivasan04} and \citet{Pauwels2002} in using SKU market shares as weights.  Prices are not deflated because there is strong evidence that people are sensitive to nominal rather than real price changes  \citep{Shafir1997} over short time periods.

\medskip

We use data from January 1993 to July 1994, 77 weeks in total. We neither use 
data before 1993 since they contain missing observations, nor
observations after 1994 since \citet{Srinivasan04}  pointed out that manufacturers made extensive use of `pay-for-performance' price promotions as of 1994, which are not fully reflected in the Dominick's database. This data range is short relative to the dimension of the VAR, which calls for a regularization approach such as the Sparse VAR. For all stores, we collect data on sales, promotion and pricing  for all 17 categories. Only for cigarettes, no promotion variable is included in the VAR since none of the SKUs in that category were promoted during the observation period.

We estimate a separate VAR model for each store, which allows to evaluate the robustness of the findings. The multivariate time series entering the VAR model are the log-differenced sales  ($\mathbf{Y_t}$), differenced promotion ($\mathbf{M_t}$), and log-differenced prices  ($\mathbf{P_t}$).\footnote{Following standard practice, we first test for stationarity. A stationarity test of all individual time series using the Augmented Dickey-Fuller test indicates that most time series in levels are integrated of order 1.}  The dimensions of the time series are represented in Table \ref{data}.  We use the Vector Autoregressive model, with endogenous promotion and prices,
\begin{equation}\label{eq: application model}
\left[ \begin{matrix} \mathbf{Y_t} \\ \mathbf{P_t} \\ \mathbf{M_t}  \end{matrix}  \right] =
B_0 + B_1 \left[ \begin{matrix} \mathbf{Y_{t-1}} \\ \mathbf{P_{t-1}} \\ \mathbf{M_{t-1}}  \end{matrix}  \right]
+ \ldots + B_p \left[ \begin{matrix} \mathbf{Y_{t-p}} \\ \mathbf{P_{t-p}} \\ \mathbf{M_{t-p}}  \end{matrix}  \right] + \mathbf{e_t}.
\end{equation}
Averaged across stores, the selected value of $p$ is two for the Sparse VAR.
Also for the Bayesian estimators, the lag order of the VAR is selected using the BIC criterion, which is one for the majority of the stores. 

\linespread{1.2}
\begin{table}
\begin{center}
\caption{Description of the 15 data sets. Each data set contains multivariate time series for sales ($\textbf{Y}_{t}$), promotion ($\textbf{M}_{t}$) and prices ($\textbf{P}_{t}$). \label{data}}
\small
\begin{tabular}{cccccc} \hline
\rule{0pt}{3ex} Store & Number of & \multicolumn{3}{c}{Dimension} & \\
 & Time Points  &  $\textbf{Y}_{t}$ & $\textbf{M}_{t}$ & $\textbf{P}_{t}$ & Total \\ \hline
Store 1-15 \rule{0pt}{3ex} & 77 & 17 & 16 & 17 & 50\\ \hline
\end{tabular} \\
\end{center}
\end{table}
\linespread{1.5}
%--------------------------------------------------------------------------------------------
\section{Empirical Results}
%--------------------------------------------------------------------------------------------

We focus on the effects of prices, promotions and sales in category A on the sales (or demand) in category B, where A and B belong to the product category network.  We first study the direct effects.   
For instance, there is no direct effect of price of A on  sales of B if the corresponding estimated regression coefficients  are equal to zero at all lags.   
Then we turn to the complete chain of direct and indirect effects using Impulse Response Functions.  
For instance, price in category A indirectly influences sales in category B when the price of category A influences the price, promotion or sales in a certain other category C which, in turn, influences the sales of category B. Since we work in a time series setting, both direct and indirect effects are dynamic in the sense that the effect occurs with a certain delay.

\subsection{A network of product categories}
\begin{figure}
	\begin{center}
		\includegraphics[width=9cm, angle=-90]{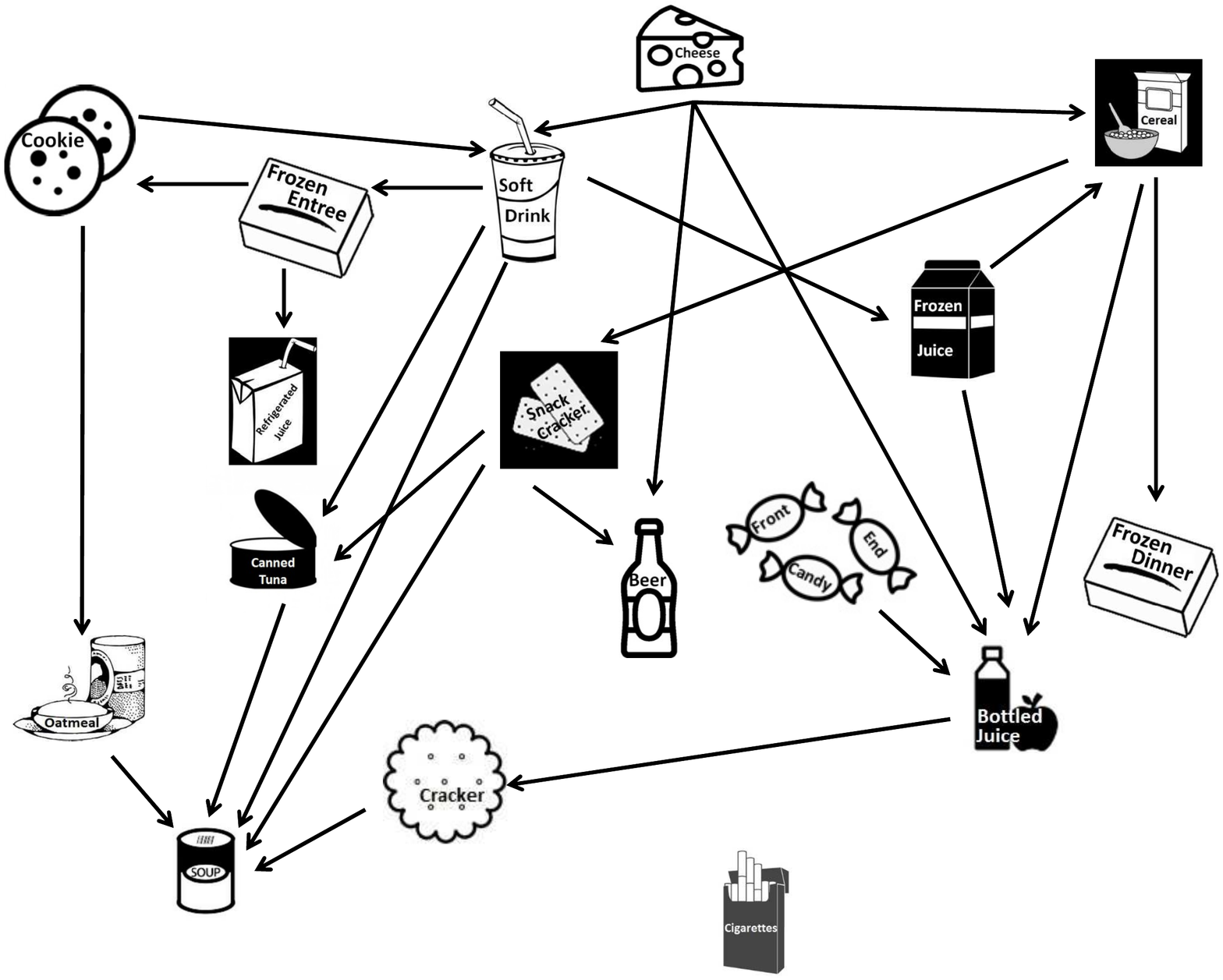}
	\end{center}
	\caption{Cross-category effect network of prices on sales: a directed edge is drawn from one category to another if its price influences sales in the other category for the majority of stores. \label{crosscatPrice}} 
\end{figure}

\begin{figure}
	\begin{center}
		\includegraphics[width=8.5cm, angle=-90]{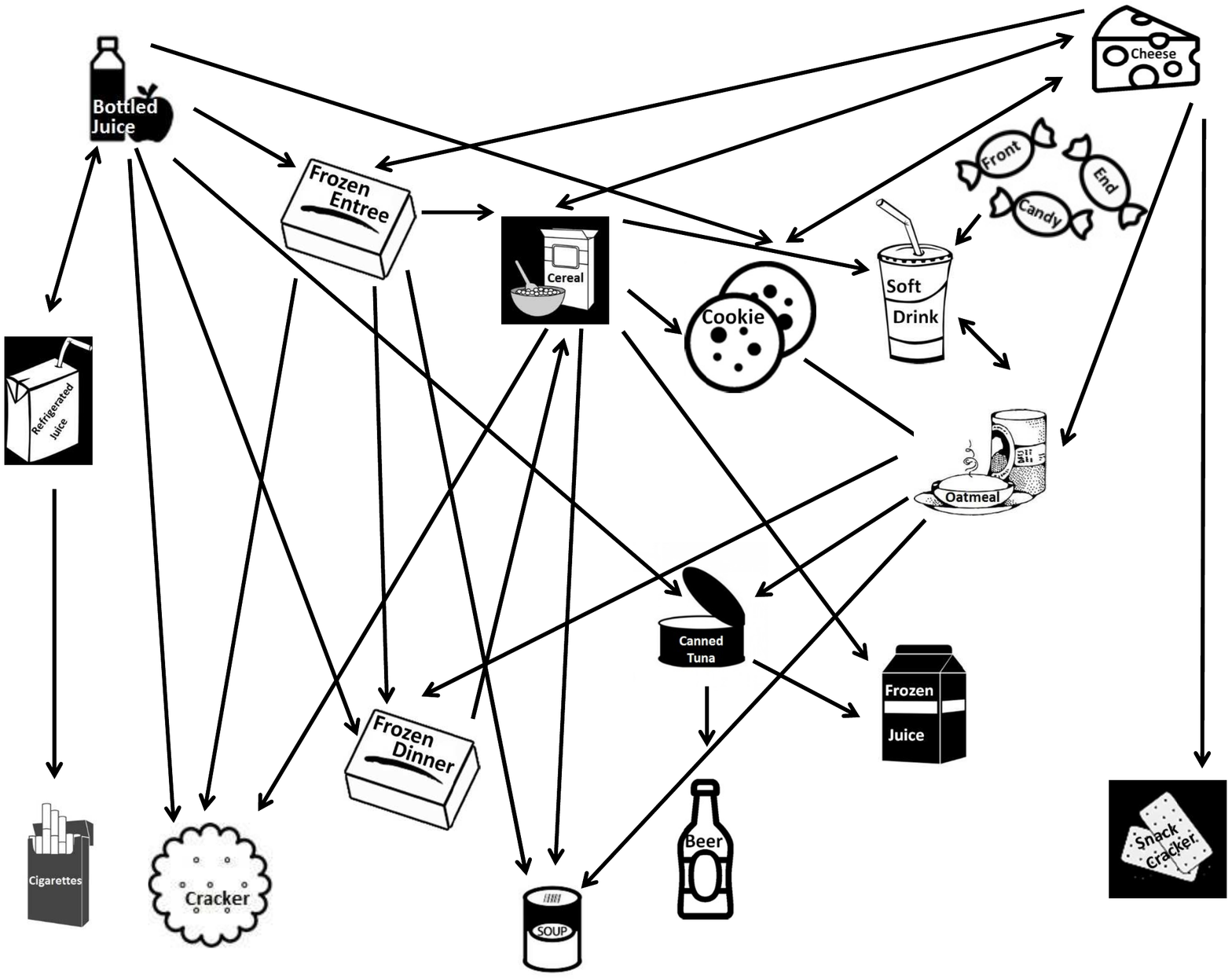}
	\end{center}
	\caption{Cross-category effect network of promotions on sales: a directed edge is drawn from one category to another if its promotion influences sales in the other category for the majority of stores. \label{crosscatPromo}} 
\end{figure}

\begin{figure}
	\begin{center}
		\includegraphics[width=8.5cm, angle=-90]{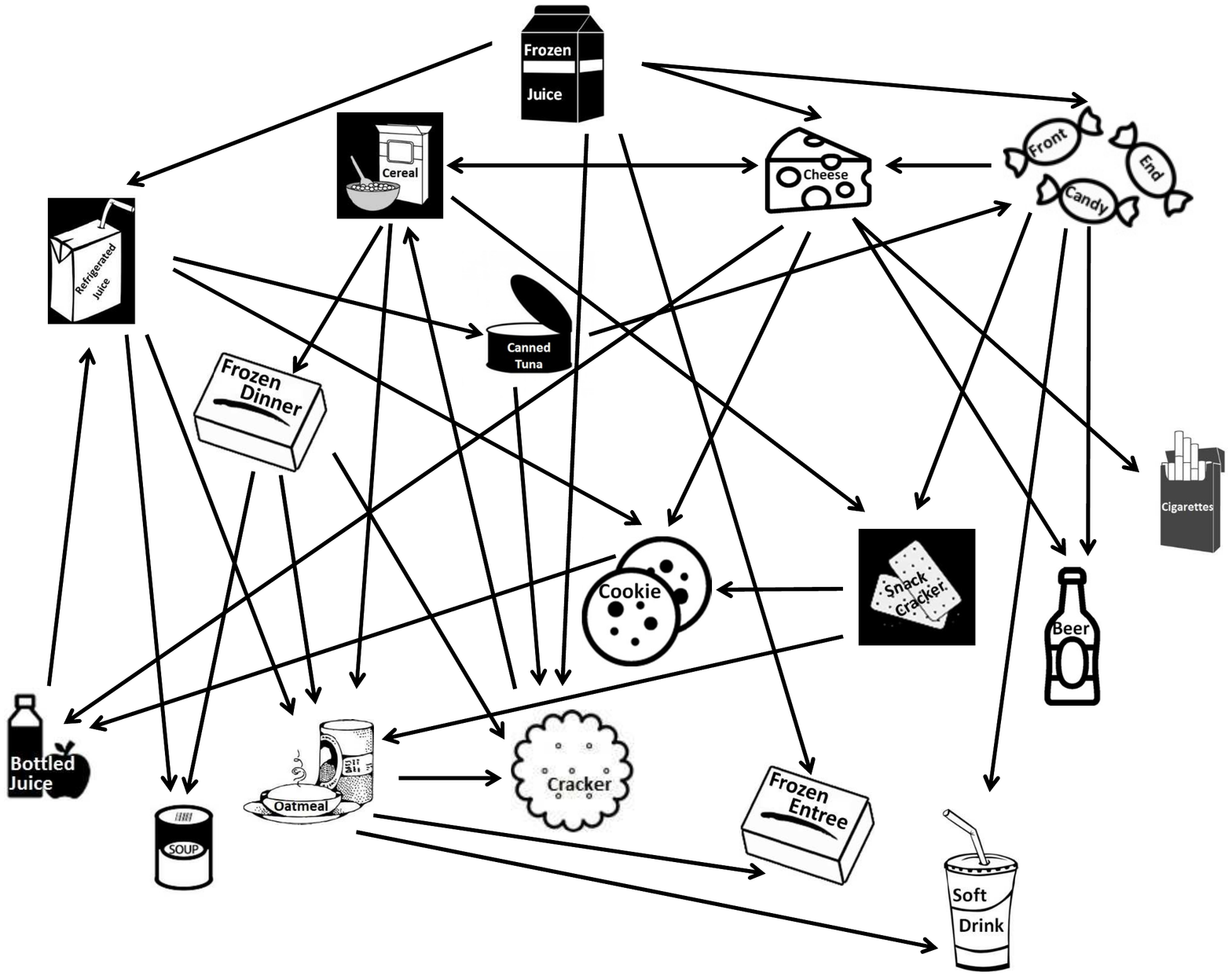}
	\end{center}
	\caption{Cross-category effect network of sales on sales: a directed edge is drawn from one category to another if its sales influences sales in the other category for the majority of stores.\label{crosscatSLS}} 
\end{figure}

%Network
We analyze cross-category demand effects as a network of interlinked product categories of which prices, promotions and sales in one category have an effect on sales in other categories. Recently, network perspectives have been increasingly used by marketing researchers to model, for example, the network value of a product in a product network \citep{Singer2013} or to investigate the flow of influence in a social network \citep{Zubcsek11}. In our case, the 17 product categories are the nodes of the network. We estimate the Sparse VAR for 15 stores separately. If the Sparse VAR estimation results indicate, by giving a non-zero estimate, that prices in one category have a direct influence on sales in another category in the majority of the 15 stores, a directed edge is drawn between them. The resulting directed network is plotted in Figure \ref{crosscatPrice}. Similarly, Figures \ref{crosscatPromo} and \ref{crosscatSLS} present cross-category effects of respectively  promotion and sales on sales. If promotion or sales in one category directly influence sales in another category, respectively, this is indicated by a directed edge.

\linespread{1.2}
\begin{table}
\centering
\caption{Proportion of nonzero within and cross-category effects of price, promotion and sales on sales, averaged across 15 stores and 17 product categories.} \label{Within-Cross}
\begin{tabular}{lccccccccc}
  \hline
 &&& Price &&& Promotion &&& Sales   \\
  \hline
Within-category &&& 34\% &&& 30\%  &&& 96\%    \\
Cross-category &&& 19\%  &&& 21\%  &&&  21\%   \\
   \hline 
\end{tabular}
\end{table}
\linespread{1.5}

%Summary Findings
A first important finding is that the cross-category networks are sparse -- not each category influences each and every other category.  While the sparse VAR estimation favors zero-effects, it does not enforce them. Here, as many as 78\% of all estimated effects are zero-effects. Table \ref{Within-Cross} summarizes the prevalence of within-and cross-category effects.  As expected, within-category effects are more common than cross-category effects. For all categories, past values of the own category's sales are selected for almost all stores. Cross-category effects of price on sales (19\%), promotion on sales (21\%) and sales on sales (21\%)  are about equally prevalent.

%Responsiveness and Influence
Next, we focus on category influence and responsiveness in the cross-category network, measured by the number of edges originating from and pointing to a category respectively.
As discussed in Section 2, destination categories are expected to be more influential, while   convenience categories are expected to be more responsive. 
  We discuss which types of categories we find to be most influential and/or responsive in the cross-category networks of prices on sales, promotion on sales, and sales on sales. 

The most influential categories in the cross-category network of prices on sales are destination categories such as Soft Drinks and Cheeses (cfr.\ each four outgoing edges in Figure \ref{crosscatPrice}). This is consistent with our expectations, as Soft Drinks is known to be a destination category \citep{Briesch2013,Shankar2014,Blattberg95}. Soft Drinks is ranked first and Cheeses third in terms of food and drink expenditures (see Table \ref{Categories}) and are both heavily promoted by retailers.  A price change in either of these categories thus strongly influences the budget constraint, which in turn influences purchase decisions in other categories.  In the cross-category network of promotions on sales, Cereals is the most influential category (cfr. five outgoing edges in Figure \ref{crosscatPromo}). \cite{Briesch2013} identified Cereals as highly ranked among the destination categories.  This is not surprising as cereals are part of daily consumption patterns and are ranked second in terms of food and drink expenditures.  In the cross-category effects network of sales on sales in Figure \ref{crosscatSLS}, we identify again Cheeses as the most influential category.  

We find convenience categories to be highly responsive to changes in other categories.
The most prominent price effects are observed for Canned Soup (cfr.\ five incoming edges in Figure \ref{crosscatPrice});
the most prominent promotion effects for Frozen Dinners, Crackers and Canned Soup (cfr. each three incoming edges in Figure \ref{crosscatPromo}); 
and the most prominent sales effects for Oatmeal and Crackers (cfr.\ each four incoming edges in Figure \ref{crosscatSLS}). 
%In line with our conceptual framework, 
These categories are typically bought out of convenience, such as Frozen Dinners and Canned Soup; or bought on occasion, such as Oatmeal and Crackers, counting for a very small percentage of food and drink expenditures (see Table \ref{Categories}). 

Routine categories such as Bottled Juices, Refrigerated Juices, Frozen Juices and Cookies score moderate-to-high on category influence but are also responsive. 
This is in line with our expectation of many grocery categories being routine categories that are moderately influential and moderately responsive. Finally, the cigarettes category is least responsive and least influential.  This finding is not surprising as cigarettes are addictive, hence, smokers probably have a stable consumption unrelated to food and drinks.
%people who smoke will buy cigarettes no matter what.

To confirm the robustness of the results obtained by Sparse VAR, we check whether category responsiveness and influence are consistent across stores. We compute Kendall's coefficient of concordance $W$  for category influence and responsiveness calculated from the graphs in Figures 2-4 at the store level.   As $W$ increases from 0 to 1, there is stronger consistency across stores.  Table \ref{KendallW} indicates that all values of Kendall's $W$ are significant.

\linespread{1.2}
\begin{table}
\begin{center}
\caption{Kendall's coefficient of concordance across stores of cross-category effects of price, promotion and sales on sales for both category influence and responsiveness. $P$-values are indicated between parentheses. \label{KendallW}}
\begin{tabular}{lccccccccc}
\hline

  &&& Price &&& Promotion &&& Sales  \\ \hline
  Influence 		&&& $\underset{(< 0.001)}{0.40}$ &&& $\underset{(< 0.001)}{0.56}$ &&& $\underset{(< 0.001)}{0.30}$    \\
 Responsiveness \rule{0pt}{3ex}	&&& $\underset{(<0.001)}{0.30}$ &&& $\underset{(0.001)}{0.16}$ &&&  $\underset{(< 0.001)}{0.17}$  \\\hline
\end{tabular}
\end{center}
\end{table}
\linespread{1.5}

\subsection{Impulse Response Functions}
% Impulse Response Functions: effect sizes and signs
For each store, we estimate the Sparse VAR and compute the corresponding Impulse Response Functions (IRFs). The effect size of an impulse is obtained by summing the absolute values of the responses across the first 10 lags of the IRF, where we take absolute values in order not to average out positive and negative response. We compute effect sizes of impulses in price, promotion or sales in one product category on the sales in the same (within) category or another (cross) category. In Table 6, we report the within and cross-category price, promotion and sales effect sizes, averaged across the 15 stores and the  product categories.

Table \ref{Effectsizes} indicates that, for example, a one standard deviation price shock leads to an accumulated absolute change of .004 in own sales growth over a time period of 10 lags. As for the direct effects, we systematically find that within-category effects are larger in magnitude than cross-category effects, especially for sales and prices. For the marketing mix, promotions exert stronger within- as well as cross-category effects than price changes.

\linespread{1.2}
\begin{table}
\centering
\caption{Size of within and cross-category effects of price, promotion and sales on sales, summed across 10 lags of the IRF, averaged across stores and product categories, and in absolute value.} \label{Effectsizes}
\begin{tabular}{lccccccccc}
  \hline
 &&& Price &&& Promotion &&& Sales    \\
  \hline
Within-category &&& 0.004 &&& 0.006 &&& 0.057   \\
Cross-category &&& 0.002 &&& 0.005 &&& 0.002    \\ \hline
\end{tabular}
\end{table}
\linespread{1.5}

\begin{table}
\caption{Cross-category price, promotion and sales effects on sales summed across 10 lags of IRFs and averaged across stores. We present only the five largest positive and negative effects. \label{Crosscateffects}}
\footnotesize
\resizebox{0.95\textwidth}{!}{\begin{minipage}{\textwidth}
\centering
\begin{tabular}{lllcc|llllcc}
\hline
\multicolumn{6}{l}{\textbf{Cross-category price effects}} & \multicolumn{5}{l}{\textbf{ }}\\ \hline
Price && Sales   && Effect && Price && Sales   && Effect  \\  
impulse &&  response  &  &   && impulse &&  response  &  &   \\ \hline
 \multicolumn{4}{c}{\underline{Perceived complements}} &&&  \multicolumn{4}{c}{\underline{Perceived substitutes}} &\\
Soft Drinks && Canned Tuna   && -0.0209 && Front-end-candies && Bottled Juices  && 0.0120\\
Soft Drinks  && Frozen Entrees  &&-0.0182&& Soft Drinks  && Frozen Juices   && 0.0060\\
Canned Tuna  &&  Canned Soup   &&  -0.0173 && Snack Crackers && Beer  && 0.0058\\
Cereals && Frozen Dinners  &&  -0.0104 && Cookies && Oatmeal && 0.0056\\
Bottled Juices && Crackers && -0.0074 && Frozen Juices && Bottled Juices && 0.0023\\ \hline
\multicolumn{6}{l}{\textbf{Cross-category promotion effects}} & \multicolumn{5}{l}{\textbf{ }}\\  \hline
Promotion && Sales  &&  Effect && Promotion && Sales  &&  Effect  \\ 
impulse &&  response &&   && impulse &&  response  &  &   \\ \hline
Bottled Juices && Frozen Entrees &&  0.0586  && Oatmeal && Canned Tuna && -0.0214\\
Cheeses && Frozen Entrees &&  0.0421 && Cheeses && Cookies &&  -0.0160 \\
Crackers && Frozen Entrees &&  0.0246 && Bottles Juices && Canned Tuna && -0.0158  \\
Frozen Dinners && Frozen Entrees &&  0.0170&& Refrigerated Juices && Canned Tuna && -0.0128 \\
Snack Crackers && Frozen Entrees &&  0.0127 && Cereals && Cheeses &&  -0.0127 \\ \hline

\multicolumn{6}{l}{\textbf{Cross-category sales effects}} & \multicolumn{5}{l}{\textbf{ }}\\   \hline
Sales && Sales  &&  Effect && Sales && Sales  && Effect  \\
impulse &&  response  &  &  && impulse &&  response &&   \\  \hline
Front-end-candies && Soft Drinks && 0.0191 && Snack Crackers  && Oatmeal  && -0.0154 \\
Oatmeal && Frozen Entrees &&  0.0123 && Frozen Juices && Frozen Entrees && -0.0120 \\
Canned Tuna && Crackers && 0.0094  && Cereals  && Frozen Dinners  && -0.0099 \\
Front-end-candies  && Beer  && 0.0086 && Snack Crackers &&  Cookies &&  -0.0087\\
Snack Crackers && Frozen Dinners && 0.0064 && Refrigerated Juices  && Canned Tuna  &&-0.0084  \\\hline
\end{tabular}
\end{minipage} }
\end{table}

To get more insight in the sign of the cross-category effects, we summarize each IRF by the sum of the first 10 responses, and average this number over all stores. Table \ref{Crosscateffects} reports the five  largest positive and negative cross-category effects of price, promotion and sales on sales.

%Discussion of cross-category effects, linking back to section on cross-category management
\medskip

%Cross-category price effects
\textit{Cross-category price effects.}  We investigate whether consumers perceive categories as complements or as substitutes. Complementary and substitution effects occur between categories because they are consumed together or separately.  Following the standard economic definition \citep{Pashigian98}, complements are defined as goods having a negative cross-price elasticity, whereas substitutes are defined as goods having a positive cross-price elasticity. 
 We find evidence of two important drivers of cross-category price effects: consumption relatedness and the budget constraint. 
 
\begin{figure}
\begin{center}
\includegraphics[width=14.1cm]{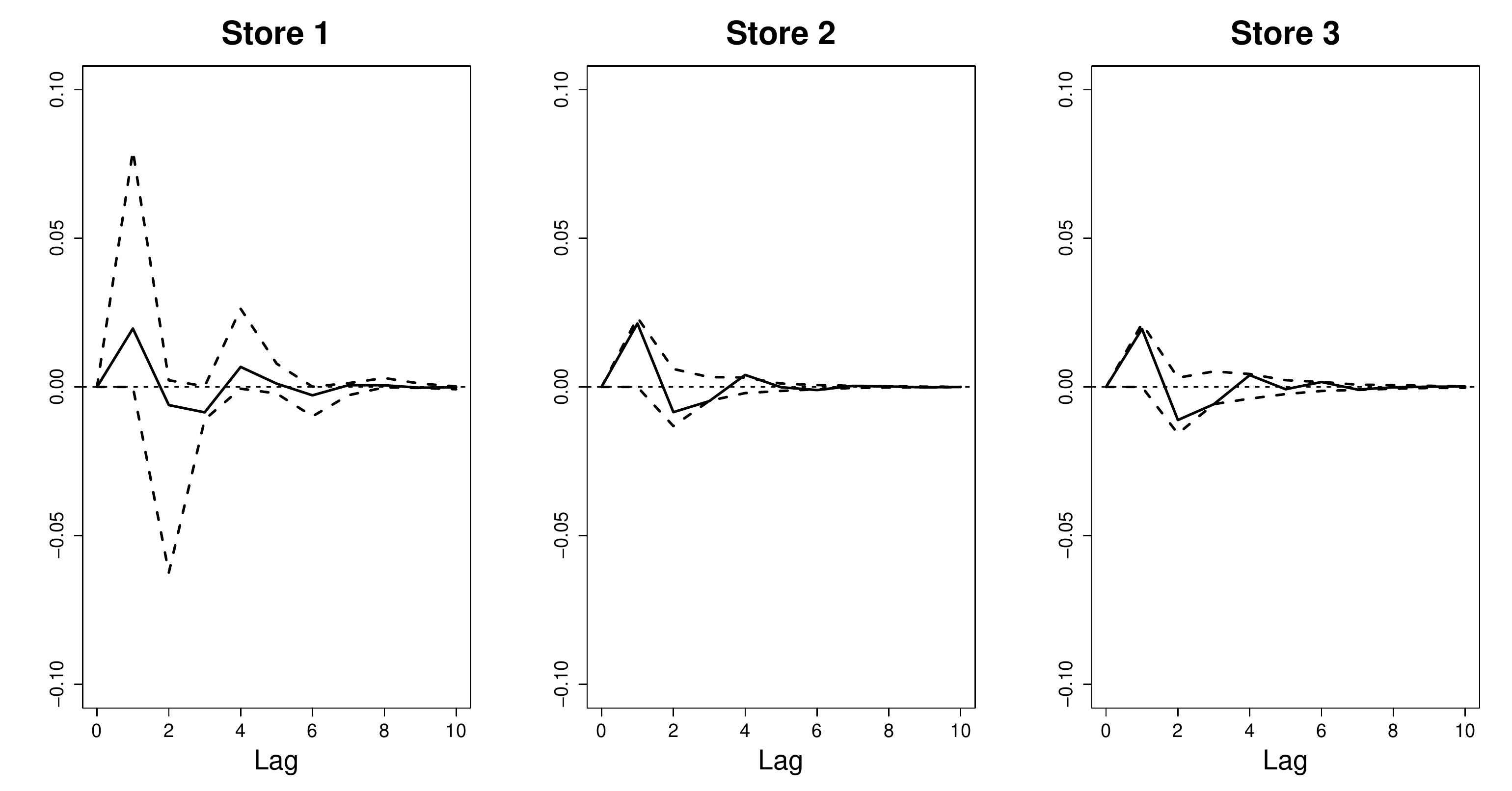}
\end{center}
\caption{Impulse response function: response of frozen juices sales growth to a one standard deviation impulse in the price of soft drinks. \label{IRF_PRICE}}
\end{figure}

%Affinity in consumption
As an example of consumption relatedness, consider Soft Drink prices and Frozen Juices.  An increase in Soft Drink prices makes consumers spend more on other drinks as a compensation, in particular Frozen Juices (see Table \ref{Crosscateffects}). The joint dynamic effect of a one standard deviation price impulse of Soft Drinks on the sales response growth of Frozen Juices is depicted in Figure \ref{IRF_PRICE} for the first three stores in the data set. Note that the instantaneous effect is estimated as exactly zero since the Sparse VAR puts the corresponding effect in the $\widehat\Sigma$ matrix to zero. We see a sharp increase in Frozen Juices sales growth one week after the soft drink price increase, indicating substitution.  However, the next two weeks, sales growth of Frozen Juices slows down, which could indicate stockpiling behavior \citep{Gangwar13}.

Another example of consumption relatedness is Soft Drinks and Frozen Entrees.  As can be seen from Table \ref{Crosscateffects}, we find a strong negative effect of Soft Drink prices on Frozen Entrees.  This might be due to the fact that Soft Drinks and Frozen Entrees are consumed together. We do not find the opposite effect of price changes in Frozen Entrees on the sales of Soft Drinks. This asymmetry arises because Soft Drinks is a destination category (high influence), while Frozen Entrees is a convenience category (highly responsiveness).

%Budget constraint
Concerning the budget constraint, prominent cross-category price effects are observed for Soft Drinks and Cereals, both destination categories.  Soft Drinks and Cereals account for a relatively large proportion of the expenditures of US families (respectively 22\% and 14\% of spending on food and drinks, see Table \ref{Categories}), which indicates that the budget constraint is an important source of cross-category effects. 

\medskip

%Cross-category promotion effects
\textit{Cross-category promotion effects.}
The results in Table \ref{Crosscateffects} indicate that branding and promotion intensity are important drivers of cross-category promotion effects.
%Branding: information see Dominick manual
Concerning branding, cross-category promotion effects are observed for categories that share brands such as Frozen Dinners and Frozen Entrees (e.g. the frozen prepared foods brand ``Stouffer's").
Concerning promotion intensity, prominent cross-category promotion effects are observed for categories in which a high percentage of the SKUs is promoted, such as Cheeses and Bottled Juices (respectively 28\% and 26\% of SKUs, on average, are promoted in our data.) A promotion impulse in such categories might either trigger join consumption (e.g. Bottled Juices and Frozen Entrees),  or deter consumption (e.g. Cheeses and Cookies).

\medskip

%Cross-category sales effects
\textit{Cross-category sales effects.} 
In Table \ref{Crosscateffects}, we find evidence of two important drivers of cross-category effects of sales on sales: affinity in consumption and the budget constraint. 
%Affinity in consumption
Prominent cross-category sales effects occur because of affinity in consumption. Some categories are jointly consumed  towards a common goal, such as Front-end-candies and Soft Drinks/Beer  (for a light meal); while others such as Snack Crackers and Cookies are purchased as replacements since consumers might perceive them to have a similar functionality.
%Budget constraint
Concerning the budget constraint, we find some cross-category sales effects between seemingly unrelated categories such as Refrigerated Juices and Canned Tuna.

\medskip

Importantly, the results from Table \ref{Crosscateffects} are in line with our findings on category influence and responsiveness. Destination categories such as Soft Drinks, Cereals and Cheeses mainly influence sales in other categories through their price, promotion or sales impulses. Convenience categories such as Frozen Entrees and Frozen Dinners are more responsive to changes in other categories.  Routine categories, such as Cookies, are moderately influential and moderately responsive, while occasional categories, such as Oatmeal, are highly responsive.

\subsection{Robustness checks}
\textit{Alternative penalty function.} We investigate the robustness of the results to the choice of the penalty function. We re-estimate the models using the Sparse VAR with elastic net instead of the grouplasso penalty (a short explanation of the elastic net is given in Section 4). The managerial insights obtained by Sparse VAR with either grouplasso or elastic net are very similar. 
Similarities are that 
(i) within-category effects are more common and larger in magnitude than cross-category effects, 
(ii) destination categories such as Cheeses and Cereals are very influential,
(iii) convenience categories such as Frozen Entrees, and occasional categories such as Crackers are very responsive
(iv) routine categories such as Bottled Juices, Refrigerated Juices and Cookies are both influential and responsive
(v) the most prominent cross-category  effects of price, promotion and sales on sales are highly overlapping.

\medskip

\textit{Alternative data period.} We also check the performance of the Sparse VAR on the post-1994 data. Retailers made extensive use of ``pay-for-performance" price promotions that are not fully reflected in the Dominick's database. The data generating process might have changed in this period. Therefore, we should not assume constant parameter values. We re-estimate the model on the post-1994 data (data from October 1995 until May 1997) and verify its performance. In the post-1994 period, similar conclusions can be drawn with respect to within versus cross-category effects and category influence and responsiveness.
Some differences are observed in the post-1994 period concerning the impulse response functions. These differences occur due to an altered strategy concerning average pricing and promotion intensity in the 17 product categories in the post-1994 period compared to the 1993-1994 period. Detailed results are available from the authors upon request.

\medskip

\textit{Alternative sparsity parameter selection.}  Our results are based on the BIC to select the penalty parameters.  We also ran the analysis using AIC as a selection criterion for the penalty function.  While the model selected by AIC are slightly less sparse, the substantive insights do not change. 

%--------------------------------------------------------------------------------------------
\subsection{Forecast Performance}
%--------------------------------------------------------------------------------------------
Although prediction is not the main goal of the proposed methodology, we deem it important to show that the Sparse VAR can compete with other methods in terms of prediction accuracy. We estimate model \eqref{eq: application model} for each store and perform a forecast exercise (cfr. Section 4), using a rolling window of length $S=67$. One-step-ahead forecasts of sales for each product category  are computed for $t=S,\ldots,T-1$, with $T=77$. The same estimation methods as in Section 4 are used.

Results on the sales predictions are summarized in Table \ref{Forecasts} by the Mean Absolute Forecast Error (MAFE), averaged across time and over the 17 product categories and 15 stores.  The MAFE should be seen as a measure of forecast accuracy, not as a measure of managerial relevance of the obtained results.
The variable selection methods Sparse VAR, 1-step and Iterative Restricted LS perform, on average, better than the methods that don't perform variable selection. This indicates that sparsity improves prediction accuracy. Sparse VAR and Iterative Restricted LS achieve the best forecasting performance. A Diebold-Mariano test \citep{Diebold95} confirms that latter two methods significantly outperform the other methods. We conclude that the improvement in interpretability of the model obtained by Sparse VAR, as discussed in the previous section, does not come at the cost of lower forecast performance.

\linespread{1.2}
\begin{table}
\begin{center}
\caption{Mean Absolute Forecast Error (MAFE) for category-specific sales, averaged over the 15 stores and the 17 product categories. $P$-values of a Diebold-Mariano test comparing the Sparse VAR to its alternatives are indicated between parentheses. \label{Forecasts}}
\small
\begin{tabular}{lcccccc}
  \hline
   \rule{0pt}{3ex} & Sparse VAR & & \multicolumn{2}{c}{Restricted LS} &  \multicolumn{2}{c}{Bayesian Methods}     \\
  \rule{0pt}{3ex} &   & LS  & 1-step  & Iterative & Minnesota & NIW     \\
  \hline
MAFE & 736.80 &  $\underset{(<0.01)}{1298.54}$   & $\underset{(<0.01)}{784.96}$   & $\underset{(0.38)}{734.82}$ & $\underset{(<0.01)}{875.47}$ & $\underset{(<0.01)}{1078.03}$     \\ \hline
\end{tabular}
\end{center}
\end{table}
\linespread{1.5}

%%%%%%%%%%%%%%%%%%%%%%%%%%%%%%%%%%%%%%%%%%%%%%%%%%%%%%%%%%%%%%%%%%%%%%%%%%%%%%%%%%%%%%%%%
\section{Discussion}
%%%%%%%%%%%%%%%%%%%%%%%%%%%%%%%%%%%%%%%%%%%%%%%%%%%%%%%%%%%%%%%%%%%%%%%%%%%%%%%%%%%%%%%%%

% Main contribution
This paper presents a Sparse VAR methodology to detect the inter-relationships in a large product category network.  In the cross-category demand effects application, we detect an important number of cross-category demand effects for a large number of categories. We find that categories have asymmetric roles: While destination categories are more influential, convenience categories are more responsive.
We identify main perceived cross-category effects but also detect cross-category effects between categories that are not directly related at first sight. Hence, the need to study -- potentially a large number of  -- product categories simultaneously. While cross-category effects are prevalent, many of them are still absent, calling for a sparse estimation procedure that succeeds in highlighting the main inter-relationships in the product category network.

% Aggregate store level data
We identify category influence and responsiveness in our cross-category demand effects application using aggregate store level data. Other cross-category studies, such as \cite{Russell97,Ainslie98,Russell99,Russell00,Elrod02} use market basket data. Since the availability and use of such market basket data pose difficulties to managers, they rarely use market basket data for category analysis \citep{Shankar2014}. As managerial decisions are often made at the category level, managers prefer to work with more readily available aggregate store level data. Hence, using aggregate category store level is managerially relevant \citep{ailawadi:09, leeflang:12}.

% Limitation
A first limitation of our approach is that we use aggregate category data, which might lead to biased estimates when there is heterogeneity on the SKU level \citep{Dekimpe00}. Second, our model does not allow to estimate cross-category effects on the individual consumer level. Insights into the behavior of consumers are revealed using market basket data, which requires a very different modeling approach. Despite these limitations, aggregate category data are highly relevant from the perspective of category management within the store.

% Advantages of Sparse VAR
An important advantage of the Sparse VAR is that it overcomes the dimensionality problem -- it results in a parsimonious model with minimal structural constraints.  We show that this leads to more accurate estimation and prediction results as compared to standard Least Squares methods.
If the researcher wishes to restrict some of the parameters to zero a priori, using marketing theory, this is of course still possible to implement with the Sparse VAR. The same holds for the reverse, i.e.\ forcing some variables to be included in the model, which can be done by adjusting the penalty on the regression coefficients in \eqref{mincrit}.

% Scope of the methodology
The methodology presented in this paper is relevant in a variety of other settings. First, Sparse VAR  can be used to study competitive demand effects across many competitors. The VAR is ideal for measuring competitive effects since it is able to capture own- and cross-elasticity of sales to both pricing and marketing spending \citep{Srinivasan04, Horvath08}.
Typically only three competitors are included  in such studies, while using the Sparse VAR allows for a much larger number to be included. Second, in the field of international marketing research there is an increased interest in studying cross-country spill-over effects, as for example in \cite{Albuquerque07}, \cite{VanEverdingen09} and \cite{Kumar02}. Every country that is added to the data set leads to an increase in the number of cross-country parameters to be estimated. Using the proposed methodology, a large VAR model could be built which allows spill-over effects between many countries.  Finally, the Market Response Model could be extended with data on online word of mouth or online search, which are now readily available. Especially in the Big Data era, most companies collect an abundance of variables \citep{BigData2013}, such that large VAR models will become even larger as more granular data become available. 

\bigskip \noindent
{\bf Acknowledgments.} The authors thank the Editors, Shankar Ganesan and Murali K. Mantrala, and two anonymous referees for their valuable comments that have improved the paper significantly. Financial support from the FWO (Research Foundation Flanders) is also gratefully acknowledged (FWO, contract number 11N9913N).

\begin{appendices}
\numberwithin{equation}{section}
\section{Penalized Likelihood Estimation}
\noindent
We iteratively solve the minimization problem \eqref{mincrit} for $\beta$ conditional on $\Omega$ and then for $\Omega$ conditional on $\beta$.

\medskip \noindent
{\it Solving for $\beta|\Omega$:}  When $\Omega$ is fixed, the minimization problem in \eqref{mincrit} is equivalent to minimizing
\begin{equation}\label{mincritbeta}
\hat{\beta}|\Omega = \underset{\beta}{\operatorname{argmin}} \frac{1}{n} (\tilde{y}-\tilde{X} \beta)^{\prime} (\tilde{y}-\tilde{X} \beta) + \lambda_1 \sum_{g=1}^{G} ||\beta_g||_2 \, ,
\end{equation}
where $\tilde{y}= Py$, $\tilde{X}=PX$, and $P$ is a matrix such that $P^{\prime}P=\tilde{\Omega}$. 
The transformation of the data to $\tilde{y}$ and $\tilde{X}$ ensures that the resulting model has uncorrelated and homoscedastic error terms. The above minimization problem is convex if $\Omega$ is nonnegative definite. The minimization problem is equivalent to the groupwise lasso of \cite{Yuan06}, implemented in the R package \verb+grplasso+ \citep{Rgrplasso}.

\noindent
{\it Solving for $\Omega|\beta$:} When  $\beta$ is fixed, the minimization problem in  \eqref{mincrit} reduces to
\begin{equation}\label{mincritOmega}
\hat{\Omega}|\beta = \underset{\Omega}{\operatorname{argmin}} \, \frac{1}{n} (y-X \beta)^{\prime} \tilde{\Omega} (y-X \beta)- \log|\Omega| + \lambda_2 \sum_{k \neq k'} |\Omega_{kk'}| \, ,
\end{equation}
which corresponds to penalized covariance estimation. Using  the glasso algorithm
of \cite{Friedman07}, available in the R package \verb+glasso+ \citep{Rglasso},  the optimization problem in \eqref{mincritOmega}
is solved.

We start the algorithm by taking $\widehat{\Omega}=I_q$ and iterate until convergence. We iterate until $max_s |\hat{\beta}_{s,i}-\hat{\beta}_{s,i-1}|<\epsilon$, with $\hat{\beta}_{s,i}$ the $s^{th}$ parameter estimate in iteration $i$ (same for $\hat{\Omega}$) and the tolerance $\epsilon$ set to $10^{-3}$. 

\paragraph{Selecting the Sparsity Parameters and the order of the VAR}
We first determine the optimal values of $\lambda_1$ and $\lambda_2$ for a fixed value of $p$, the order of the VAR. The sparsity parameters $\lambda_1$ and $\lambda_2$ are selected according to a minimal Bayes Information Criterion (BIC).
In the iteration step where $\beta$ is estimated conditional on $\Omega$,  we solve  \eqref{mincritbeta} over a range of values for $\lambda_1$ and select the one with lowest value of
\begin{equation}\label{eq: BICbeta}
BIC_{\lambda_1} = -2 \log L_{\lambda_1} + k_{\lambda_1} \log(n),
\end{equation}
where $L_{\lambda_1}$ is the estimated likelihood, corresponding to the first term in \eqref{mincritbeta},  using sparsity parameter $\lambda_1$. Furthermore, $k_{\lambda_1}$ is the number of non-zero estimated regression coefficients  and $n$ the number of observations.
Similarly, for selecting $\lambda_2$, we use the BIC given by
\begin{equation}\label{eq: BIComega}
BIC_{\lambda_2} = -2 \log L_{\lambda_2} + k_{\lambda_2} \log(n) \, .
\end{equation}
Finally, we select the order $p$ of the VAR. We estimate the VAR for different values of $p$. The optimal values of  $\lambda_1$ and $\lambda_2$ are determined for a each of those values of $p$. We select the order $p$ of the VAR  using BIC:
\begin{equation}\label{eq: BICp}
BIC_{(p, \lambda_1(p), \lambda_2(p))} = -2 \log L_{(p, \lambda_1(p), \lambda_2(p))} + k_{(p, \lambda_1(p), \lambda_2(p))} \log(n) \, , 
\end{equation}
where $L_{(p, \lambda_1(p), \lambda_2(p))}$ and $k_{(p, \lambda_1(p), \lambda_2(p))}$ depend on the value $p$ and the optimally chosen values of $\lambda_1(p)$ and $\lambda_2(p)$ for that specific value of $p$.
\end{appendices}

\linespread{1}
\bibliographystyle{asa}
\bibliography{bibfile}
\end{document}